\documentclass[conference]{IEEEtran}
\IEEEoverridecommandlockouts
\usepackage{hyperref}
\usepackage{cite}
\usepackage{amsmath,amssymb,amsfonts}
\usepackage{algorithmic}
\usepackage{graphicx}
\usepackage{siunitx,etoolbox}
\usepackage{textcomp}
\usepackage{xcolor}
\usepackage{pifont}
\newcommand{\xmark}{\ding{55}}  %  cross mark
\newcommand{\cmark}{\ding{51}}  %  check mark (optional)

\def\BibTeX{{\rm B\kern-.05em{\sc i\kern-.025em b}\kern-.08em
    T\kern-.1667em\lower.7ex\hbox{E}\kern-.125emX}}
\begin{document}

\title{Fear Moves Markets: Sentiment-Augmented POMP for Volatility Modeling of Bitcoin Returns\\

\thanks{Code and reproducibility resources: \url{https://github.com/abeyankargiridharan/Novel_Approach_to_Volatility_Analysis_on_Bitcoin_Returns}}
}
\author{\IEEEauthorblockN{Abeyankar Giridharan}
\IEEEauthorblockA{\textit{Dept. of Statistics} \\
\textit{University of Michigan}\\
Ann Arbor, USA \\
abeygiri@umich.edu}
\and
\IEEEauthorblockN{Chang Li}
\IEEEauthorblockA{\textit{Dept. of IOE} \\
\textit{University of Michigan}\\
Ann Arbor, USA \\
liichang@umich.edu}
\and
\IEEEauthorblockN{Suvrorup Mukherjee}
\IEEEauthorblockA{\textit{Dept. of Statistics} \\
\textit{University of Michigan}\\
Ann Arbor, USA \\
suvrom@umich.edu}
\and
\IEEEauthorblockN{Xinhe Wu}
\IEEEauthorblockA{\textit{Dept. of Statistics} \\
\textit{University of Michigan}\\
Ann Arbor, USA \\
xinhwu@umich.edu}
}

\maketitle

\begin{abstract}
Cryptocurrency markets exhibit extreme price swings and sentiment-driven regime shifts, which traditional volatility models often fail to capture. To address this, we develop a partially observed Markov processes (POMP) model augmented with sentiment and heavy-tailed distributions. Specifically, we extend Bretó’s framework by including the Fear and Greed Index (FGI) as an exogenous regressor in the latent volatility dynamics and replacing Gaussian measurement noise with a Student’s t distribution. We fit the model via simulation-based inference using daily Bitcoin returns and FGI data from January 2020 to April 2025. Our approach significantly outperforms three benchmark models in log-likelihood and filter stability. Moreover, our model yields interpretable parameters consistent with known features of financial volatility. These results demonstrate that incorporating sentiment signals and heavy-tailed noise improves the modeling of volatility and regime shifts in cryptocurrency markets.
\end{abstract}


% \textcolor{red}{\section{Uniform Terminology}}
% \textcolor{red}{plug-and-play =  simulation-based}

% \textcolor{red}{particle filtering = SMC}
% \begin{itemize}
%     \item explain local search/ global seach
%     \item explain metric: ESS and cond. loglik
%     \item explain filter diagnostic plot; paramter convergence plot
% \end{itemize}

\section{Introduction} \label{sec:intro}
% Volatility analysis has been a key area of statistical research, with models such as Heston's Stochastic volatility model (HSV), Generalized Autoregressive Conditional Heteroskedasticity (GARCH) models, and hidden Markov processes frameworks developed to capture the complex dynamics of financial markets. However, the emergence of new markets for cryptocurrencies characterized by sentiment-driven behavior and extreme volatility calls for novel approaches to better capture their patterns.

% The Fear and Greed Index (FGI) ranges from 0 to 100, where lower values indicate heightened market fear and higher values indicate increased investor greed. It is derived from market momentum, social media sentiment, and trading volume to quantify investor psychology. Given the sentiment-driven nature of cryptocurrency markets like Bitcoin, the FGI could provide invaluable information to better capture regime shifts and volatility patterns that traditional models might overlook.

Volatility modeling plays a key role in risk management, derivative pricing, and portfolio optimization. Traditional approaches such as Heston’s stochastic volatility model (HSV) \cite{HSV}, GARCH, and hidden Markov models (HMMs) are well suited to equity and fixed-income markets but often fail to capture the extreme swings and sentiment-driven regime shifts that characterize cryptocurrencies. As Bitcoin and similar assets attract retail and institutional investors alike, understanding how fear and greed influence price fluctuations has become increasingly important to academic researchers and market participants.

In this paper, we address the challenge of modeling cryptocurrency volatility characterized by both heavy-tailed returns and rapidly shifting investor sentiment. 
We leverage the Fear and Greed Index (FGI) in our model, which captures market psychology using momentum, trading volume, and social media data. FGI ranges from 0 to 100, where lower values indicate heightened market fear and higher values reflect increased investor greed. 

Our key contributions are twofold:
\begin{itemize}
    \item \textbf{Sentiment-Augmented POMP}: We extend the POMP volatility model proposed by Bretó \cite{Bret__2014} by introducing the FGI as an exogenous regressor in the latent volatility dynamics. Consequently, we enable the model to adapt to shifts in market psychology.
    \item \textbf{Heavy-Tailed Measurement Noise}: We replace the Gaussian observation noise with a Student’s t distribution, allowing the model to accommodate the fat tails and occasional price spikes endemic to crypto returns.
\end{itemize}

We compare this enhanced model with three benchmarks, including GARCH, the HSV model, and the original Bretó model. Our results show that our enhanced POMP model consistently outperforms all benchmarks using log-likelihood scores as a measure of model fit, demonstrating improved capacity to capture volatility clustering and sentiment-driven regime shifts.

\section{RELATED WORK} \label{sec:related_work}
Recent literature about modeling cryptocurrency volatility  \cite{vol-lit} can be categorized into three categories.

The first is \textit{Traditional Econometric Models}.
GARCH-type models remain foundational in volatility modeling. For instance, Chong et al. \cite{chong1999performance} evaluated multiple GARCH variants on the Kuala Lumpur Stock Exchange (KLSE) index and found that the exponential GARCH (EGARCH) model was particularly effective in capturing asymmetries and improving out-of-sample forecasting accuracy. These insights inform our selection of GARCH as a baseline model for Bitcoin volatility.

The second is \textit{Machine Learning and Hybrid Approaches}.
Recent studies have explored the integration of statistical learning techniques such as support vector machines (SVMs), neural networks, and ensemble methods with volatility models. These hybrid models often enhance predictive accuracy and flexibility in high-frequency or nonstationary environments, particularly relevant to cryptocurrency markets.

% \begin{figure*}[!t]
%     \includegraphics[width=\textwidth]{bit_fgi_comb_manual.png}
%     {\centering
%         \caption{Daily Bitcoin closing prices (Jan 2020 – Apr 2025)}\label{fig:bit_fgi_comb}
%     }
% \end{figure*}

The third is \textit{State-Space and Regime-Switching Models}.
The use of Hidden Markov Models (HMMs) and POMPs has gained traction for capturing latent volatility dynamics and structural breaks. Eisler et al. \cite{eisler2007volatility} modeled price returns and volatility using a two-dimensional diffusion system under an expOU framework which uncovered lognormal volatility behavior and power-law relationships with trading volume. Augustyniak et al. \cite{augustyniak2019new} extended this with a factorial hidden Markov volatility model that incorporates regime shifts, jumps, and leverage effects, demonstrating strong empirical performance across time horizons.

Our work falls within this third category and builds upon the state-space framework proposed by Bretó \cite{Bret__2014}, which introduces a random-walk leverage mechanism to allow flexible, time-varying asymmetric volatility dynamics. In contrast to prior work, we augment the POMP framework with sentiment-based features (i.e., FGI) to capture behavioral responses to market sentiment. This decision is motivated by recent work from Singhal \cite{singhal2023role}, which highlights the influence of investor psychology on volatility during bearish regimes and economic downturns.

\section{Data Description and Preprocessing} \label{sec:data}

Our analysis uses daily Bitcoin closing prices and the Fear and Greed Index (FGI) over the period from January 2020 to April 2025. This window captures several major macroeconomic and market-specific events, including the COVID-19 shock, the 2021 and 2024 bull runs, and the 2022 crypto market downturn. These events offer a diverse range of volatility regimes, therefore the period is particularly suitable for studying dynamic risk behavior in cryptocurrency markets. Furthermore, using daily data is essential to capture short-term volatility clustering and sentiment-driven regime shifts that lower-frequency series may fail to reflect.

\subsubsection{Bitcoin Returns}
\begin{figure}[!t]
    \centering
    \includegraphics[width=\columnwidth, height=0.47\textheight]{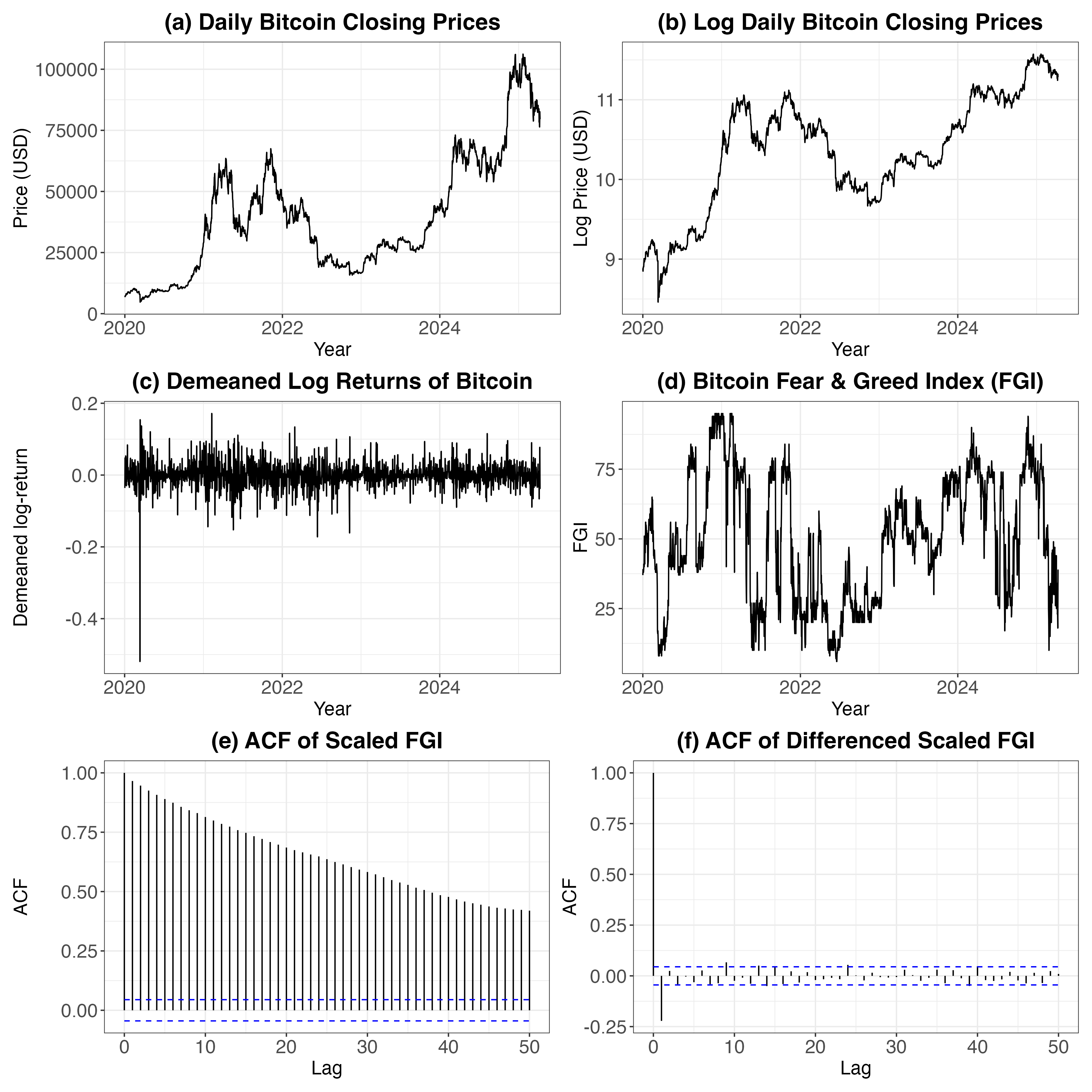}
    \caption{Exploratory visualization and preprocessing of Bitcoin price and sentiment data from January 2020 to April 2025.}
    \label{fig:bit_fgi_comb}
\end{figure}

Our data preprocessing follows the common practice in finance to work with a zero‐mean, centered log‐return series.
Let $Z_n$ denote the raw closing prices at time $n$, where $n=1,\ldots,N$.
We first compute the continuously compounded return
\[ A_{n+1} = \log (Z_{n+1}) - \log(Z_{n}),\]
then we get a centered series (i.e., our demanded log returns)
\begin{align} \label{eq:demanded_return} 
    Y_n &= A_n - \bar{A},
\end{align}
where $\bar{A}$ is the sample mean of $\{ A_n\}$. This transformation removes mean drift and helps stabilize the variance, thereby making the series suitable for volatility modeling where the zero-mean assumption is essential. 
The demeaned log returns $Y_n$ in Fig \ref{fig:bit_fgi_comb}(c) reveals persistent large volatility during periods of market stress such as mid-2022 and lasting small volatility during consolidation phases of 2023. These patterns are also known as volatility clustering.
In contrast, the raw Bitcoin series in Fig. \ref{fig:bit_fgi_comb}(a) obscures these clustered dynamics. Therefore, we show the necessity to obtain demeaned log returns. 
% Visual inspection of the demeaned log returns $Y_n$ reveals heightened volatility during periods of market stress such as mid-2022 and relative calm during consolidation phases. 
% , supporting the presence of volatility clustering in the data. These patterns are visually evident in Fig.~\ref{fig:bit_fgi_comb} (a) and (b), which show the raw price series $Z_n$ and the demeaned log returns $Y_n$, respectively.

\subsubsection{Fear and Greed Index} 
 As introduced earlier, the Fear and Greed Index (FGI) captures investor sentiment using a composite of market momentum, volume, volatility, and social signals. Over the 2020–2025 period, the index exhibits sharp fluctuations that correspond closely with major market events. A prominent peak appears in early 2021 and indicates extreme greed. This peak coincides with Bitcoin’s surge fueled by institutional investments such as Tesla’s \$1.5 billion Bitcoin purchase and MicroStrategy’s continued accumulation. In contrast, a deep trough in mid-2022 reflects heightened fear following the collapse of the Terra/Luna ecosystem, which triggered widespread liquidations and a broader crisis of confidence in the crypto sector. These sentiment swings illustrate the strong influence of macroeconomic shocks, corporate actions, and systemic events on investor psychology, thus motivates the inclusion of the FGI as a behavioral covariate in our volatility modeling.

The raw FGI exhibits strong serial dependence, as indicated by the slow decay in autocorrelations shown in Fig. \ref{fig:bit_fgi_comb}(e). To prepare the series for modeling, we first center it around its range midpoint (i.e., 50) and then apply first-order differencing. This transformation reduces autocorrelation and yields a stationary series, as confirmed by the sharp decay in the differenced ACF shown in Fig. \ref{fig:bit_fgi_comb}(f).
Centering around 50 preserves interpretability: positive values of the transformed series correspond to greed-dominated sentiment, while negative values indicate fear. The rationale for enforcing stationarity and the role of the transformed FGI within the POMP model are discussed further in Section \ref{sec:evaluation}.

\section{Methodology} \label{sec:method}

\subsection{The Original Bretó Model} \label{sec:Bretó}

We first review the original Bretó model. This state-space model captures stochastic volatility and incorporates a dynamic leverage effect by modeling time-varying volatility through a latent log-volatility process. The observed returns are assumed to be conditionally Gaussian with variance driven by this latent process. At time $n$, the observed return $Y_n$ is modeled as:
\begin{align}
    Y_n = \exp\left(H_n/2\right) \varepsilon_n,
\end{align}
where $H_n$ is the log-volatility and $\varepsilon_n \sim \mathcal{N}(0,1)$.
The model includes two latent states $H_n$ and $G_n$.
$H_n$ is structured as an AR(1) process with an additional leverage effect term: 
\begin{align} \label{eq:Bretó_hn}
    H_n &= \mu_h(1 - \phi) + \phi H_{n-1} \notag \\
    \quad & \quad + \beta_{n-1} R_n \exp\left(-H_{n-1}/2\right) + \omega_n,
\end{align}
where $\beta_n$ is shock scale, $R_n$ is transformed random walk and $w_n$ is process noise. They are defined as below:
\begin{align*}
    \beta_n  &=  Y_n \, \sigma_\eta \sqrt{1 - \phi^2}, \\
    R_n      &=  \tanh(G_n) =\frac{\exp{(2G_n)} - 1}{\exp{(2G_{n})} +1} \in (-1, 1) , \\
    \omega_n &\sim \mathcal{N}\left(0, \, \sigma_\eta^2 (1 - \phi^2)(1 - R_n^2)\right).
\end{align*}

The second latent state $G_n$ is a random walk
\begin{align}
    G_n = G_{n-1} + \nu_n,
\end{align}
where $\nu_n \sim \mathcal{N}(0, \sigma_\nu^2)$.

Intuitively, the model captures leverage effects through the interaction between $\beta_{n}$ and $R_n$ (i.e., the first term in the second line of Eq. \eqref{eq:Bretó_hn}). 
Specifically, a negative return $Y_n$ combined with a negative $R_n$ results in a positive contribution to the log-volatility $H_n$
thereby increasing the conditional variance of $Y_n$ (i.e., $\exp (H_n)$).
In contrast, large positive returns are less frequent in practice, and the magnitude of positive $R_n$ values tends to be smaller due to the properties of the transformation $\tanh$. As a result, the effect on volatility in such cases is more moderate. This asymmetric volatility response to positive versus negative shocks constitutes the leverage effect.
The state-space model structure is suitable for likelihood-based inference via particle filtering.

% \subsection{The Bretó Model and Its Enhancements} \label{sec:mod-Bretó}

% We build upon the partially observed Markov processes (POMP) volatility model originally proposed by Bretó (2014), which models time-varying volatility through a latent state process and incorporates a dynamic leverage effect. The observed returns $\{Y_n\}$ are modeled as:
% \[
% Y_n = \exp\left(\frac{H_n}{2}\right) \varepsilon_n, \quad \varepsilon_n \sim \mathcal{N}(0,1)
% \]
% The log-volatility $\{H_n\}$ evolves as:
% \[
% H_n = \mu_h(1 - \phi) + \phi H_{n-1} + \beta_{n-1} R_n \exp\left(-H_{n-1}/2\right) + \omega_n
% \]
% where $R_n$ captures leverage effects and is defined in terms of an auxiliary latent variable $G_n$ as:
% \[
% G_n = G_{n-1} + \nu_n, \quad 
% R_n = \frac{\exp(2G_n) - 1}{\exp(2G_n) + 1}
% \]
% The latent state vector at time $n$ is thus $X_n = \{G_n, H_n\}$, and the model jointly captures both volatility dynamics and asymmetry via $R_n$. Parameters are estimated using particle filtering for likelihood evaluation and iterated filtering (MIF2) for parameter optimization.

% \vspace{0.5em}
% \textbf{Incorporating Sentiment and Heavy Tails.} 

\subsection{The Enhanced Bretó Model} \label{sec:mod-Bretó}

% \subsubsection{Incorporating Sentiment and Heavy Tails.} 
We enhance the original Bretó model through two key modifications to reflect the behavioral and statistical characteristics of cryptocurrency markets. First, we introduce the Fear and Greed Index (FGI) as an exogenous regressor into the volatility process. Since the raw FGI series exhibits strong persistence and non-stationarity, we apply first-order differencing to obtain a stationary series (Fig.~\ref{fig:bit_fgi_comb}(f)). This transformation is necessary because the differenced FGI enters directly into the latent volatility dynamics. Including a non-stationary regressor in the state equation can induce unstable and non-stationary latent dynamics, and may even lead to explosive behavior when combined with high autoregressive persistence. By differencing and scaling the FGI, we ensure that it captures short-term sentiment fluctuations while preserving model stability, identifiability, and interpretability.
% Since the raw FGI is non-stationary, we center it around 50 and apply first-order differencing to obtain a stationary series:
% \[
% \Delta \text{FGI}_{\text{scaled},n} = \frac{\text{FGI}_n - 50}{50} - \frac{\text{FGI}_{n-1} - 50}{50}
% \]
% As shown in Fig. \ref{fig:bit_fgi_comb}, the original FGI series exhibits strong autocorrelation, while the transformed series shows no significant autocorrelation, that is, we have stationarity  after differencing.

% \begin{figure}[!t]
%     \centering
%     \includegraphics[width=\columnwidth]{bit_fgi_comb_tt.png}
%     \caption{Autocorrelation of Transformed FGI}
%     \label{fig:bit_fgi_comb_t}
% \end{figure}

This allows us to modify the volatility evolution equation as:
\begin{align} \label{eq:Bretó_enhance}
    H_n &= \mu_h(1 - \phi) + \phi H_{n-1} + \beta_{n-1} R_n \exp\left(-H_{n-1}/2\right) \notag\\
        &\quad + \gamma \, \Delta \mathrm{FGI}_{\text{scaled},\,n-1} + \omega_n
\end{align}

The coefficient $\gamma$ captures the directional influence of sentiment on volatility. A negative $\gamma$ with decreased FGI (heightened fear) increases volatility. A positive $\gamma$ with rising FGI (increased greed) also amplifies market volatility.

Second, we replace the standard Gaussian observation noise $\varepsilon_n$ with a Student’s t-distribution to better capture the heavy-tailed nature of Bitcoin returns. This adjustment improves robustness to outliers and large shocks. After empirical testing, we find that a t-distribution with $\nu = 5$ degrees of freedom offers the best fit.

Together, these enhancements yield a sentiment-aware, heavy-tailed volatility model well-suited to the structural features of cryptocurrency markets.

\subsection{Other Baseline Models}
We also briefly review two additional baseline models, $\text{GARCH}(p,q)$ model and the HSV model. 

The $\text{GARCH}(p,q)$ captures time-varying volatility by modeling the conditional variance of returns as a deterministic function of past squared residuals and past variances. The model specification metric is based on AIC or log-likelihood. Let $Y_n$ and $h_n$ be the observed return and volatility at time $n$, respectively. The model is defined as:
\begin{align*}
Y_n &= \mu + \varepsilon_n,\\
h_n &= \omega + \sum_{i=1}^q \alpha_i \varepsilon_{t-i}^2 + \sum_{j=1}^p \beta_j h_{t-j},
\end{align*}
where $\mu$ is mean return, $\varepsilon_n = \sqrt{h_n} z_n$ and $z_n \sim \mathcal{N}(0,1)$.

The HSV model introduces a latent variance process \( V_n \) that evolves stochastically with mean reversion and innovation-driven noise. Its dynamics are specified by:
\begin{align*}
Y_n &= \sqrt{V_n} \, \varepsilon_n ,\\
V_n &= (1 - \phi)\theta + \phi V_{n-1} + \xi \sqrt{V_{n-1}} \, \omega_n ,
\end{align*}
where $\varepsilon_n \sim \mathcal{N}(0,1)$ and $\omega_n \sim \mathcal{N}(0,1)$.

In summary, GARCH enforces a relatively inflexible, purely autoregressive structure on the conditional variance, whereas state-space models (i.e., HSV and the original Bretó model) allow for latent‐state dynamics.  Neither GARCH nor HSV captures leverage effects or admits exogenous covariates. By contrast, the Bretó class of models (including our enhanced version) can accommodate both features. A concise comparison of structural form, leverage treatment, and exogenous input inclusion for all four models appears in Table \ref{tab:models}.

\begin{table}[ht]
\centering
\caption{Comparison of volatility models}
\label{tab:models}
\setlength{\tabcolsep}{12pt}
\renewcommand{\arraystretch}{1.2}
\begin{tabular}{|c|c|c|c|c|}
\hline
\textbf{Model} & \textbf{SS} & \textbf{SV} & \textbf{LE} & \textbf{SA} \\
\hline
$\text{GARCH}(p,q)$ & \xmark & \xmark & \xmark & \xmark \\
\hline
HSV model & \cmark & \cmark & \xmark & \xmark \\
\hline
Original Bretó & \cmark & \cmark & \cmark & \xmark \\
\hline
Enhanced Bretó & \cmark & \cmark & \cmark & \cmark \\
\hline
\end{tabular}\\
\vspace{0.5em}
\noindent
\textbf{Note:}  
SS = State-space Model; SV = Stochastic Volatility; LE = Leverage Effect; SA = Sentiment Awareness. 
\vspace{-1em}
\end{table}

%, implemented via the \texttt{mif2} function in the \texttt{pomp} package~\cite{pomp}
\section{Evaluation and Results} \label{sec:evaluation}
In this section, we evaluate our proposed model using the Bitcoin dataset described in Section \ref{sec:data}. Parameter estimation is performed via the iterated filtering (IF) algorithm~\cite{if2}, which applies stochastic optimization to latent-variable models by combining parameter perturbations with particle filtering across successive iterations.
The evaluation compares our model with three benchmarks: (1) a GARCH(3,1) model selected through grid search, (2) the original Bretó model, and (3) the HSV model.

% \subsection{Filter Diagnostics}

\begin{figure}[!t]
\centering
\includegraphics[width=\columnwidth]{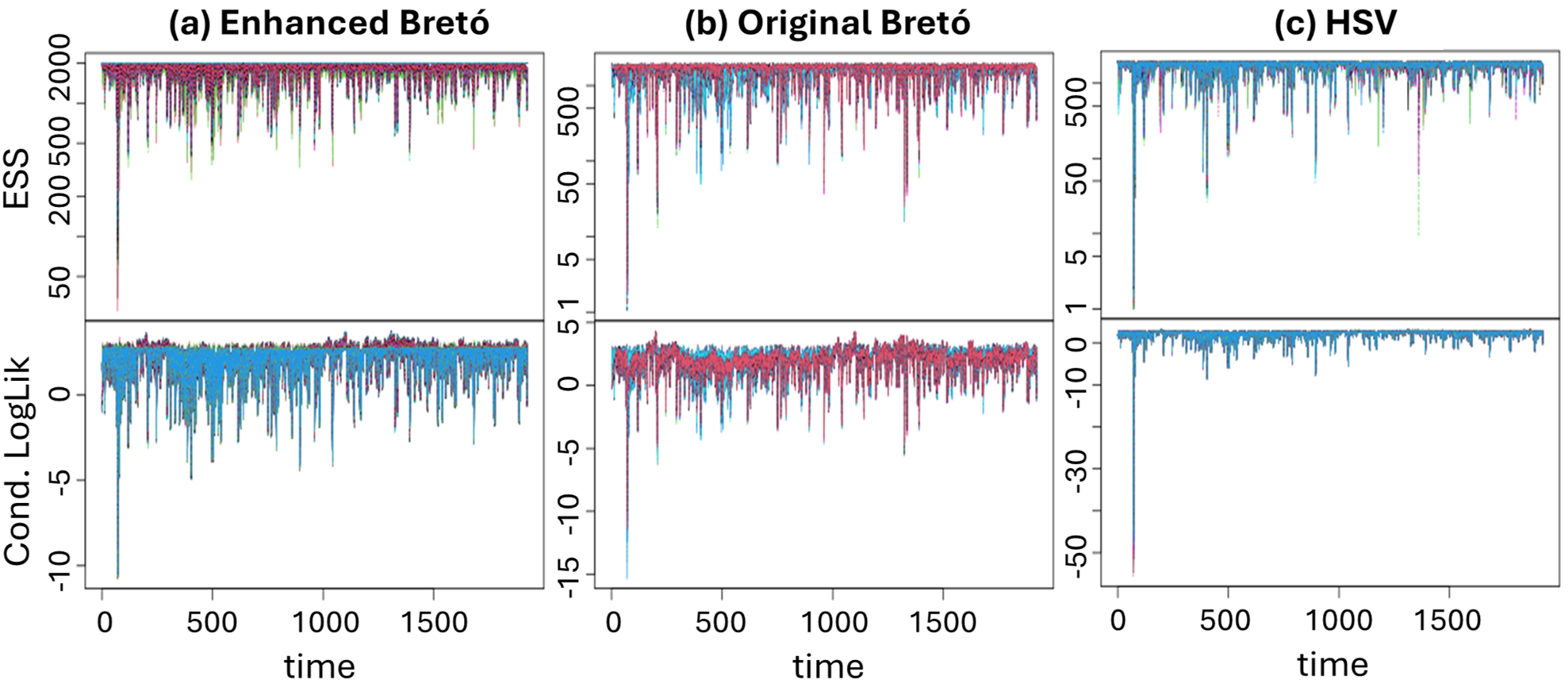}
\caption{Filter Diagnostics (Last Iteration)}\label{fig:filter_diagnostics}
\end{figure}

We first evaluate the effectiveness of our method via Effective Sample Size (ESS). ESS is defined as
\[\text{ESS}_n = \left( \sum_{i}^{N_p} w_{in}^2 \right)^{-1}, \]
where $w_{in}$ is the normalized weight of particle $i$ at time $n$ and each particle $i$ represents a simulated trajectory of the latent state.
ESS ranges from 1 (all weight on one particle, i.e., particle degeneracy) to $N_p$ (all weights equal, i.e., particle diversity). High ESS implies that the particle set remains a reliable approximation of the true posterior, while low ESS signals weight collapse and poor representation.
In our analysis, we set $N_p=2000$.
The enhanced Bretó model shows strong stability as it maintains ESS values close to the theoretical maximum (2000) across most time points in Fig. \ref{fig:filter_diagnostics}(a). 
Occasional drops below 50 reflect brief particle degeneracy, a common issue in nonlinear filtering that does not compromise overall model stability.
In contrast, both benchmark models show consistently lower ESS in Fig. \ref{fig:filter_diagnostics}(b) and (c). 
We cannot assess ESS of GARCH(3,1) since IF is not applicable to GARCH.

We further assess model performance using the conditional log-likelihood, $\log p(Y_n \mid Y_{1:n-1}, \hat{\boldsymbol{\theta}})$, under the final parameter estimate $\hat{\boldsymbol{\theta}} = (\hat{\sigma}\eta, \hat{\sigma}\nu, \hat{\gamma}, \hat{\mu}_h, \hat{G}_0, \hat{\phi}, \hat{H}_0)$, as defined in Eq.~\ref{eq:Bretó_enhance}.
Our model fits the observed data better, achieving higher conditional log-likelihood values than the benchmarks, as shown in the lower panel of Fig.~\ref{fig:filter_diagnostics}.

% \subsection{Model Convergence}
Next, we analyze parameter convergence using IF to maximize the log-likelihood. Due to the high nonlinearity of the likelihood surface, we first conduct a local search from an informed starting point to identify high-likelihood regions. Subsequently, we define a plausible parameter box and run multiple global IF searches from randomly sampled initial values within this region. The parameter set with the maximum estimated log-likelihood is selected as the final estimate.

\begin{figure}
    \centering
    \includegraphics[width=\columnwidth]{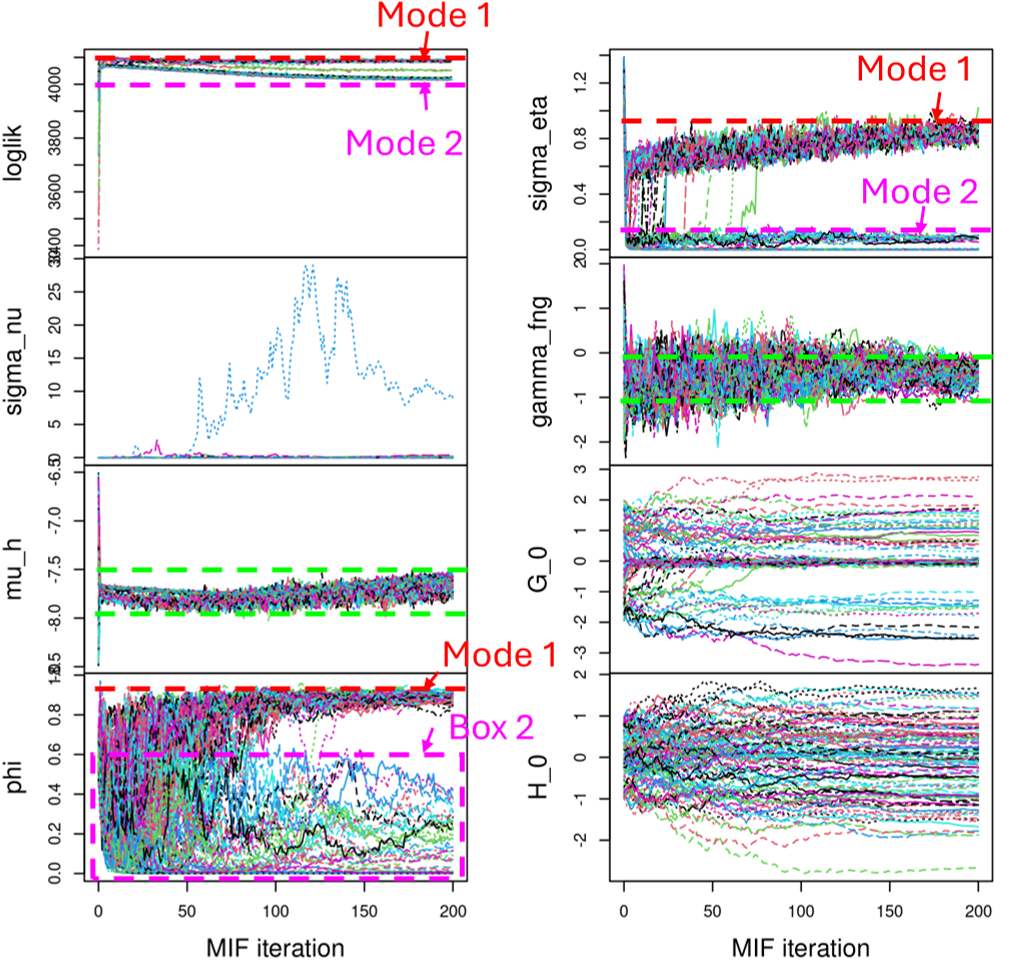}
    \caption{Convergence Plot of the enhanced Bretó Model}
    \label{fig:convg}
\end{figure}

The parameter convergence plot for the enhanced Bretó model is shown in Fig. ~\ref{fig:convg}. It overlays 100 independent iterated filtering trajectories. Each path shows the evolution of the log‐likelihood and the seven model parameters \(\sigma_{\eta}\), \(\sigma_{\nu}\), \(\gamma\), \(\mu_{h}\), \(G_{0}\), \(\phi\), and \(H_{0}\) as defined in Eq.~\eqref{eq:Bretó_enhance}. Trajectories from the same search are rendered in a distinct color and line style.
The analysis below focuses on the key parameters driving the model's behavior.

% loglik
\subsubsection{Loglikelihood}
We observe a bimodal structure in the log-likelihood trajectories for our proposed model. Mode 1 peaks at around \(4100\), about \(100\) log‐likelihood units above Mode 2 at roughly \(4000\). A similar bifurcation is evident in the original Bretó model (plot omitted for brevity), whereas the HSV model exhibits a single, unimodal likelihood surface (plot omitted for brevity). This contrast underscores the superior flexibility of Bretó‐based models in capturing the complex dynamics present in our data, compared to the HSV formulation. Moreover, our proposed model achieves the highest log-likelihood of $4093.85$ compared to the benchmark models in Table \ref{tab:results}, thus we demonstrate our approach's effectiveness.

\begin{table}[ht]
\sisetup{
  table-align-uncertainty=true,
  separate-uncertainty=true,
}
%% local redefinitions
\renewrobustcmd{\bfseries}{\fontseries{b}\selectfont}
\renewrobustcmd{\boldmath}{}

\centering
\caption{Model Performance}
\label{tab:results}
\setlength{\tabcolsep}{12pt}
\renewcommand{\arraystretch}{1.2}
\begin{tabular}{|c|c|}
\hline
\textbf{Model} & \textbf{Max Log-Likelihood}  \\

\hline
Enhanced Bretó & $\boldmath 4093.85$ \\
\hline
Original Bretó & $4074.67$ \\
\hline
HSV model & $3959.11$ \\
\hline
$\text{GARCH}(3,1)$ & $3902.41$ \\
\hline

\end{tabular}\\

\noindent
\end{table}
\vspace{-0.5em}
\subsubsection{$\sigma_\eta$ and $\phi$ for leverage effect}
We also observe that both the shock scale (\(\sigma_{\eta}\)) and persistence (\(\phi\)) exhibit bimodal convergence. Although the secondary mode of \(\phi\) is less sharply defined, its values span roughly \(0\) to \(0.6\). For convenience, we refer to this region as “Box 2” for \(\phi\). Recall that
\[
\beta_n \;=\; Y_n\,\sigma_{\eta}\,\sqrt{1 - \phi^2},
\]
so \(\sigma_{\eta}\) and \(\phi\) jointly influence the evolution of \(H_n\). Consequently, their trajectories display a similar bimodal structure.
\begin{itemize}
    \item Mode 1: \(\phi\) converges near \(1\), making \(\sqrt{1-\phi^2}\approx 0\). To preserve the leverage effect, \(\sigma_{\eta}\) increases to a relatively large value.
    \item Mode/Box 2: \(\sigma_{\eta}\) concentrates at a smaller value, while \(\phi\) fluctuates between \(0\) and \(0.6\), yielding a moderately large \(\sqrt{1-\phi^2}\). Their product remains sufficient to balance the leverage term.
\end{itemize}
This complementary behavior confirms that the model adaptively trades off \(\sigma_{\eta}\) and \(\phi\) to capture volatility dynamics.

\subsubsection{$\gamma$ for sentiment awareness}
% sentiment awareness
The coefficient $\gamma$ converges to the negative range $(-1, 0)$. This indicates that negative shifts in sentiment (i.e., increasing fear) are associated with increased volatility. That is, fear exerts a stronger influence on volatility than greed. Therefore, the negative sign of $\gamma$ is both mathematically coherent and economically meaningful.

\subsubsection{$\mu_h$ for Model Drift}
We also observe that model drift ($\mu_h$) converges to the range of approximately $-7.5$ to $-8$ in the enhanced Bretó model (see Fig.~\ref{fig:convg}), while it does not exhibit any convergence pattern in the original Bretó model (plot omitted for brevity). One plausible explanation is that the addition of $\gamma \, \Delta \text{FGI}$ in Eq.~\eqref{eq:Bretó_enhance} removes a key confound between drift and exogenous shocks, thereby improving the identifiability of $\mu_h$ and allowing its estimated distribution to converge to a clear, well-defined mode.

\begin{figure}
    \centering
    \includegraphics[width=0.9\linewidth]{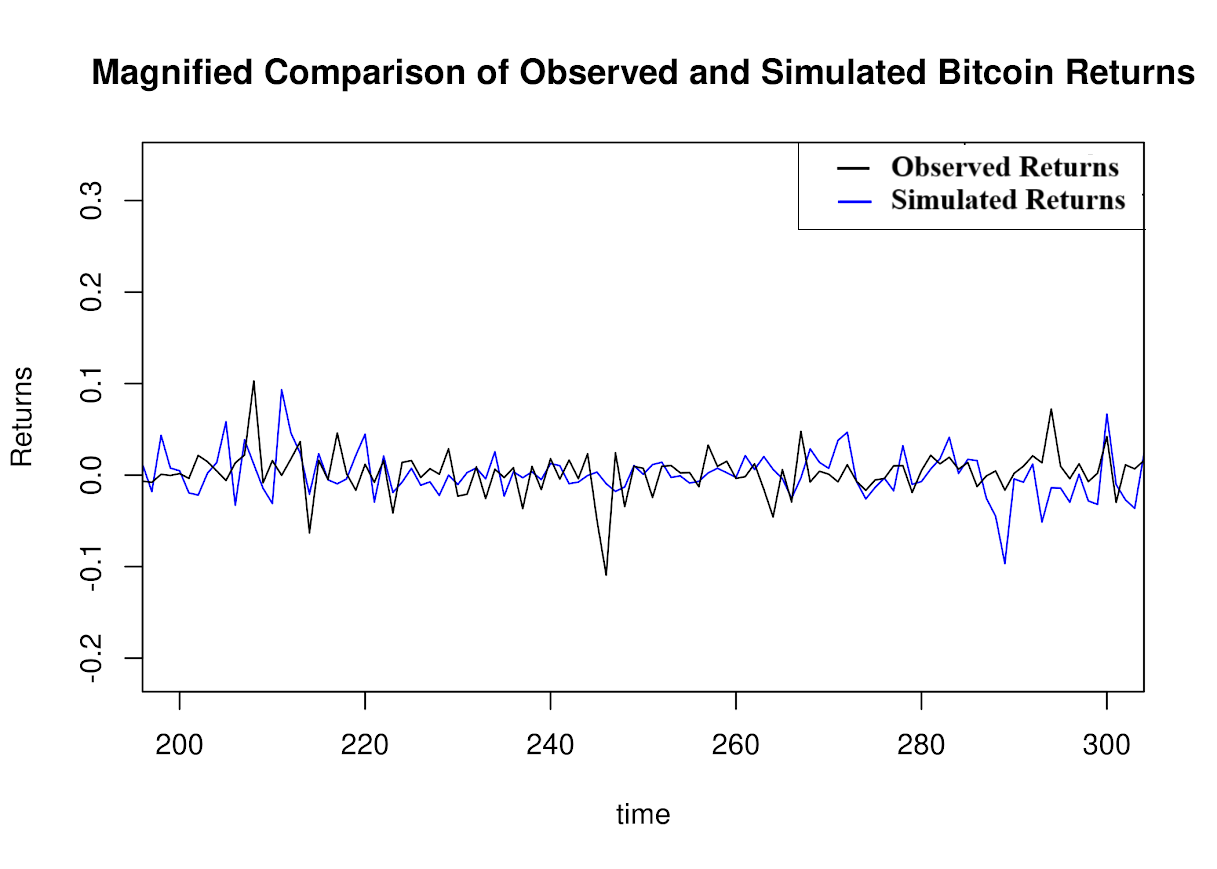}
    \vspace{-1.5em}
    \caption{Magnified comparison of observed and simulated returns.} \label{fig:simulate}
\end{figure}

% \subsection{Simulated Return Dynamics Using the Enhanced POMP Model}
In addition to interpretability and higher log-likelihood, we further demonstrate the strength of our model through simulation. We generate a sample simulation with the estimated parameters corresponding to the maximum log-likelihood of the enhanced Bretó model in Table~\ref{tab:results}. Fig.~\ref{fig:simulate} shows the simulated demeaned return $Y_n$, defined in Eq.~\eqref{eq:demanded_return}, over a randomly selected period from day 200 to day 300. The simulated curve exhibits close correspondence with the ground truth, thereby supporting the validity of our parameter estimation.

\section{Conclusions} \label{sec:conclusioin}
We propose a sentiment-aware, heavy-tailed POMP model that extends the original Bretó’s framework to address the extreme fluctuations and regime shifts in cryptocurrency markets. 
Our key innovations, incorporating FGI and implementing heavy-tailed distributions, align with established market-volatility patterns and enhance model interpretability. 
We show that our model achieves superior filter stability and the highest log-likelihood via comparisons with three benchmarks using daily Bitcoin prices and FGI series from 2020 to 2025. 
We further validate the model via simulation and confirm that simulated returns closely match observed dynamics. Notably, our results indicate that increases in fear exert a significantly stronger effect on volatility than equivalent increases in greed, which reflects the asymmetric influence of investor sentiment.

Overall, these results demonstrate that the inclusion of investor sentiment and heavy-tailed noise enhances the model's ability to capture volatility clustering and regime transitions in cryptocurrency markets. This framework also provides a foundation for extensions to other digital assets and alternative sentiment indicators.%Future work could explore extending this framework to other cryptocurrencies and investigating additional sentiment indicators.
%\begin{tcolorbox}[colback=gray!5,colframe=black!80,title=Effective Sample Size (ESS)]
%\[
%\text{ESS} = \frac{m}{1 + \text{cv}^2(W)}
%\]
%\textbf{Where:}
%\begin{itemize}
%    \item \(m\): total number of samples across all chains
%    \item \(\text{cv}^2(W)\): coefficient of variation squared of the within-chain variances
%    \item \(W\): within-chain variance
%\end{itemize}
%\end{tcolorbox}

\bibliographystyle{IEEEtran}
\bibliography{ref}

@article{vol-lit,
    author = {Almeida, José and Gonçalvesm Tiago Cruz},
    title = {A Systematic Literature Review of Volatility and Risk Management on Cryptocurrency Investment: A Methodological Point of View},
    journal = {Risks},
    year = {2022},
    doi = {10.3390/risks10050107}
}

@article{chong1999performance,
  title={Performance of {GARCH} models in forecasting stock market volatility},
  author={Chong, Choo Wei and Ahmad, Muhammad Idrees and Abdullah, Mat Yusoff},
  journal={Journal of forecasting},
  volume={18},
  number={5},
  pages={333--343},
  year={1999},
  publisher={Wiley Online Library}
}

@article{Bret__2014,
   title={On idiosyncratic stochasticity of financial leverage effects},
   volume={91},
   ISSN={0167-7152},
   url={http://dx.doi.org/10.1016/j.spl.2014.04.003},
   DOI={10.1016/j.spl.2014.04.003},
   journal={Statistics \&amp; Probability Letters},
   publisher={Elsevier BV},
   author={Bretó, Carles},
   year={2014},
   month=aug, pages={20–26} }

@inproceedings{singhal2023role,
  title={Role of greed and fear index in investment decision making},
  author={Singhal, Jyoti},
  booktitle={International conference on:'Leveraging Technology for Creating Competitive Advantage},
  year={2023}
}

@article{eisler2007volatility,
  title={Volatility: a hidden Markov process in financial time series},
  author={Eisler, Zolt{\'a}n and Perell{\'o}, Josep and Masoliver, Jaume},
  journal={Physical Review E—Statistical, Nonlinear, and Soft Matter Physics},
  volume={76},
  number={5},
  pages={056105},
  year={2007},
  publisher={APS}
}

@article{augustyniak2019new,
  title={A new approach to volatility modeling: The factorial hidden Markov volatility model},
  author={Augustyniak, Maciej and Bauwens, Luc and Dufays, Arnaud},
  journal={Journal of Business \& Economic Statistics},
  volume={37},
  number={4},
  pages={696--709},
  year={2019},
  publisher={Taylor \& Francis}
}

@article{HSV,
author = {NANCE, DEANA R. and SMITH JR., CLIFFORD W. and SMITHSON, CHARLES W.},
title = {On the Determinants of Corporate Hedging},
journal = {The Journal of Finance},
volume = {48},
number = {1},
pages = {267-284},
doi = {https://doi.org/10.1111/j.1540-6261.1993.tb04709.x},
year = {1993}
}

@article{
if2,
author = {Edward L. Ionides  and Dao Nguyen  and Yves Atchadé  and Stilian Stoev  and Aaron A. King },
title = {Inference for dynamic and latent variable models via iterated, perturbed Bayes maps},
journal = {Proceedings of the National Academy of Sciences},
volume = {112},
number = {3},
pages = {719-724},
year = {2015},
doi = {10.1073/pnas.1410597112}
}

\end{document}